\documentclass[osajnl2,twocolumn,showpacs,superscriptaddress,10pt]{revtex4-1}
\usepackage{graphicx}
\usepackage{amsmath}
\usepackage{amsfonts}%
\usepackage{amssymb}
\usepackage{mathrsfs}
\usepackage{dcolumn}% Align table columns on decimal point
\usepackage{bm}% bold math
\usepackage{float}
\usepackage{epsfig}% FOR CORRECT FIGURE CAPTIONS
\usepackage{color}
\usepackage{fancyhdr}
\begin{document}

\title{Effect of input pulse chirp on nonlinear energy deposition and plasma excitation in water}

\author{Carles Mili\'{a}n} 
\email[]{carles.milian@cpht.polytechnique.fr}
\affiliation{Centre de Physique Th\'{e}orique, CNRS, \'{E}cole Polytechnique, F-91128 Palaiseau, France}
\author{Am\'elie Jarnac}
\affiliation{Laboratoire d'Optique Appliqu\'{e}e, ENSTA ParisTech, \'{E}cole Polytechnique, CNRS, F-91761 Palaiseau, France}
\author{Yohann Brelet}
\affiliation{Laboratoire d'Optique Appliqu\'{e}e, ENSTA ParisTech, \'{E}cole Polytechnique, CNRS, F-91761 Palaiseau, France}
\author{Vytautas Jukna}
\affiliation{Centre de Physique Th\'{e}orique, CNRS, \'{E}cole Polytechnique, F-91128 Palaiseau, France}
\author{Aur\'elien Houard}
\affiliation{Laboratoire d'Optique Appliqu\'{e}e, ENSTA ParisTech, \'{E}cole Polytechnique, CNRS, F-91761 Palaiseau, France}
\author{Andr\'e Mysyrowicz}
\affiliation{Laboratoire d'Optique Appliqu\'{e}e, ENSTA ParisTech, \'{E}cole Polytechnique, CNRS, F-91761 Palaiseau, France}
\author{Arnaud Couairon}
\affiliation{Centre de Physique Th\'{e}orique, CNRS, \'{E}cole Polytechnique, F-91128 Palaiseau, France}

\begin{abstract}

We analyze numerically and experimentally the effect of the input pulse chirp on the nonlinear energy deposition from $5\ \mu$J fs-pulses at $800$ nm to water. Numerical results are also shown for pulses at $400$ nm, where linear losses are minimized, and for different focusing geometries. Input chirp is found to have a big impact on the deposited energy and on the plasma distribution around focus, thus providing a simple and effective mechanism to tune the electron density and energy deposition. We identify three relevant ways in which plasma features may be tuned.

Published article: http://www.opticsinfobase.org/josab/abstract.cfm?URI=josab-31-11-2829
\end{abstract}

\maketitle

%SEC I
\section{introduction}

Laser energy deposition in condensed dielectric media has many applications ranging from micromachining of glasses \cite{gattassNP08,UtezaASS07,ChimierPRB11} and medical laser surgery \cite{PlamannJO10,FengCPL96,VogelJASA96,VogelAPB05} to bubble formation \cite{FaccioPRE2012,SreejaLP13} and sound wave generation for oceanography \cite{JonesNRL07,FaccioPRE2012,JarnacConf13}. The initial stage of laser energy deposition consists in the generation of a localized weakly ionized plasma with typical density of one electron per hundreds of molecules in the focal region of the laser beam (see, e.g., Refs. \cite{GamalyPR2011,BulgakovaQE13} for reviews on laser-plasma interaction in solids and Refs. \cite{FengI3EJQE97,KennedyPQE97,NoackI3EJQE99,LiuAPB2002,VogelAPB05,EfimenkoJOSAB14} and references therein for works in water and other liquids).

To deposit laser energy in a well defined focal volume far from the surface, femtosecond pulses carrying energies of few $\mu$J may be used in conjunction with tight focusing geometries to avoid nonlinear effects prior to the focus. Smaller numerical apertures trigger nonlinearities at earlier stages and are closely related to the phenomenon of laser beam filamentation \cite{CouaironSPR07,JarnacPRA14}, but this allow for laser energy deposition at deeper distances from the surface (see, e.g., Refs. \cite{ArnoldAPB05,DubietisAPB06,CouaironPRE06,MinardiOL08,MinardiOL09} for investigations in water and other liquids in the last few years). 

We are interested in femtosecond laser energy deposition in water for the above-mentioned potential applications that require a well controlled localized plasma in a focal volume at depths from a few centimeters to deeper positions under the surface. From a practical point of view,  liquids provide a platform where the localized plasma tracks do not lead to permanent damage since they are naturally erased via electron-ion recombination. This allows for consecutive independent material excitations at laser repetition rates $\nu\lesssim$kHz, which are much lower than the typical hydrodynamic inverse time scales for material recovery, $\nu_{hydro}\sim$ MHz (see recent experimental results in Ref. \cite{FaccioPRE2012}).

The idea that input pulse chirp has a strong impact on nonlinear dynamics has been widely used during the last decade. For media exhibiting normal group velocity dispersion (GVD), an optimal negative input chirp makes equal the spatial focusing and temporal compression lengths, yielding enhancement of the nonlinear effects resulting in high intensities, long plasma channels and broader spectra \cite{WosteLO97,SpranglePRE02,GolubtsovQE03,ZengPRA11,DurandOE13}, generation of few-cycle \cite{HauriOE05,ParkOE08} and ultra-short \cite{VarelaOL10} pulses, and the possibility for remote spectroscopy \cite{ZengJOSAB12}. Control of input chirp has also been reported to enhance pulse collision induced spectral broadening \cite{KolesikOL07} and damage tracks in solids \cite{OndaJOSAB05,BhuyanAPL14}.% Interestingly enough, there are other approaches to enhance the generation of plasmas which involve pre-chirping Bessel \cite{PolynkinOE08} and Bessel-Gauss beams \cite{PolynkinOE09} or modify focusing conditions of chirp-free pulses \cite{EisenmannOE07}. These alternative approaches are also expected to play a relevant role in the control of energy deposition but remain, however, unexplored in the present work.

In this work, we explore numerically and experimentally the effects the input pulse chirp and focusing conditions have on nonlinear energy deposition from $\mu$J pulses at $800$ nm to water and on the electron-plasma density distribution. Comparison between numerical and experimental results for the transmission (laser energy deposition) presents a very good agreement. Numerical simulations let us acquire a deeper understanding of the spatiotemporal dynamics and have access to experimentally inaccessible data, such as the generated plasmas, fluence distributions, intensity profiles, etc. We then identify different plasma generation regimes. In particular we find three different input pulse widths that maximize different features of the plasmas. First, the minimum of the optical transmission corresponds to negatively pre-chirped pulses that generate the plasmas with the maximum possible energy. Second, further negative pre-chirping results in a plasma volume with maximized length (and still relatively high electron density), and third, plasma densities are maximized for even larger negative values of the input chirp, at the expense of the plasma channel length. We foresee that the possibility to control the density and shape of the plasmas in the focal region is promising for developing the aforementioned applications.%We stress here that some regimes have been previously explained (see, e.g., the comprehensive Ref. \cite{GolubtsovQE03}), but not all of them, in particular the third one, have been precisely identified nor interpreted before.}

% PREVIOUS VERSION OF ABOVE PARAGRAPH. In this work, we explore numerically and experimentally the effects the input pulse chirp and focusing conditions have on nonlinear energy deposition from $\mu$J pulses at $800$ nm to water and on the electron-plasma density distribution. Comparison between numerical and experimental results for the transmission (laser energy deposition) presents a good agreement. These results show that the transmission of energy from the pulse to water is maximized for an optimal value of the input pulse chirp. This is associated to the case where maximum spatial and temporal compression of the pulse are simultaneously achieved along propagation. Numerical simulations show that i) around this optimal pulse chirp, a plasma volume with maximized length (and relatively high electron density) is formed, and ii) plasma densities are maximized for larger values of the input chirp. We foresee that the possibility to control the density and shape of the plasmas in the focal region is promising for developing the aforementioned applications.

The rest of this paper is organized as follows. Section II describes the experimental setup and Section III presents the theoretical model used for numerical simulations of laser energy deposition. Section IV presents the results and shows the effects on energy deposition induced by changes in the focalization geometry and chirp. Section V is devoted to the extension of the numerical results for pulses which carrier wavelength is $400$ nm. Final conclusions and remarks are expounded in Section VI.

%SEC II
\section{Experimental setup}

\begin{figure}
\includegraphics[width=0.5\textwidth]{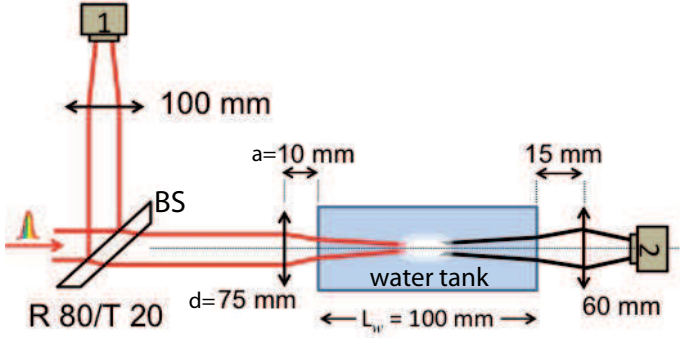}
\caption{Sketch of the experimental setup. Pulses are incident from the left. Convergent lenses with focal lengths of $100$, $75$, $60$ mm are represented by double arrows. The other elements are: power-meters $(1,2)$, beam splitter (BS), and water tank. \label{f1}}
\end{figure}

The experimental setup used for measurements and modeled below is shown in Fig. \ref{f1}. The experiment was performed by using a commercial CPA Ti:Sapphire femtosecond Laser (THALES Alpha 100) delivering $t_{pl}=45$ fs transform limited pulses with carrier vacuum wavelength $\lambda_0=800$ nm at a repetition rate of $\nu=100$ Hz. The chirped pulses are obtained by detuning the compressor stage integrated within the laser. The beam is focused inside the BK7 glass water tank with a lens of focal distance $d = 7.5$ cm in air placed at $a=1$ cm from it. The walls of the tank are $1$ cm thick and the inner length $L_w=10$ cm. Two power-meters are used to monitor simultaneously the input $(1)$ and output $(2)$ pulse energies. Their calibration with the $80\ \%$ reflectivity beam splitter (BS) provides accurate (and repeatable) measurements of the transmission as a function of input pulse chirp (see Fig. \ref{f3}) and let us keep the energy of the pulses entering the tank relatively constant: $E_{in}=5.0\pm0.1\ \mu$J. Results presented here are for distilled water, however we produced essentially identical data with tap water.

%FIG 1

%SEC III
\section{Theoretical Model}
\label{GeneralModel}

We model the propagation of the electric field envelope in Fourier space, $\widetilde{\cal E}(\omega,r,z)\equiv {\hat{\cal F}}[{\cal E}(t,r,z)]$,  along the $z$ coordinate by means of a unidirectional pulse envelope propagation equation (see, e.g., Ref. \cite{CouaironEPJST} for details),

%EQ 1
\begin{equation}
\frac{\partial \widetilde{\cal E}}{\partial z}=i\left\{{\cal K}(\omega) + \frac{\Delta_{\perp}}{2 k(\omega)} \right\}\widetilde{\cal E} +i {\cal Q}(\omega) \frac{\widetilde{{\cal P}}}{2\epsilon_0} -\frac{\widetilde{{\cal J}}}{n(\omega)2\epsilon_0 c}\label{eq1},
\end{equation}
where $k(\omega)\equiv n(\omega)\omega/c$ and $n(\omega)$ is the real valued refractive index of water (for data see Ref. \cite{MielenzAO1978}). Linear and nonlinear dispersion functions read ${\cal K}(\omega)\equiv  k(\omega)+i\beta_1/2-k_0-k'_0[\omega-\omega_0]$ and ${\cal Q}(\omega)=\omega^2/[k(\omega)c^2]$, respectively, where $k_0\equiv k(\omega_0)$, $k'_0\equiv \partial_\omega k(\omega)\vert_{\omega_0}$ are evaluated at the carrier frequency $\omega_0=2\pi c/\lambda_0$,  and $\beta_1$ is the linear attenuation coefficient. The cylindrically symmetric Laplacian $\Delta_\perp\equiv\partial_r^2+r^{-1}\partial_r$ accounts for diffraction. In the nonlinear terms of the above equation, the instantaneous Kerr nonlinearity is accounted for through the polarization $\widetilde{{\cal P}}(\omega,r,z)\equiv 2\epsilon_0n_0n_2\hat{\cal F}[I {\cal E}]$, where $n_0\equiv n(\omega_0)$, $n_2$ is the nonlinear index, and $I\equiv \epsilon_0cn_0\vert {\cal E}\vert^2/2$ the electric field intensity. The effects associated to quasi-free electrons are contained in the current $\widetilde{{\cal J}}(\omega,r,z)\equiv \widetilde{{\cal J}}_{OFI}+\widetilde{{\cal J}}_{PL}$. ${\cal{J}}_{OFI} \equiv\epsilon_0cn_0W(I)U_i[1-\rho/\rho_{nt}]{\cal E}/I$ accounts for optical field ionization (OFI), where $W(I)$ is the intensity dependent ionization rate, $U_i$ the ionization potential energy, and $\rho_{nt}$ the density of neutral molecules. The interaction of light with the plasma of electron density $\rho(t,r,z)$ is described by $\widetilde{{\cal J}}_{PL} \equiv \epsilon_0c \sigma(\omega_0) {\hat{\cal F}}[\rho {\cal{E}}]$. The cross section $\sigma(\omega)$ may be found from the Drude model as \cite{YablonovitchPRL72,CouaironEPJST,CouaironSPR07}

%TABLE
\begin{table}
\small
\resizebox{\columnwidth}{!}{\begin{tabular}{|l|c|c|c|c|}
\hline
magnitude & symbol\ (units) & \multicolumn{2}{|c|}{value} & Ref.\\
\hline\hline
vacuum wavelength & $\lambda_0\ (nm)$ & 800 & 400 & -\\
linear absorption & $\beta_1$ (cm$^{-1}$) & 0.0196 & 5 $\times$ 10$^{-4}$ & \cite{HaleAO73}  \\
MPA order & $K=\langle \frac{U_i}{\hbar \omega_0}+1\rangle$ & 5 & 3 & -\\
MPA (Keldysh) & $\beta_K$ (cm$^{2K-3}$W$^{1-K}$) & $3.6\times 10^{-50}$  & $5.4\times 10^{-24}$ & \cite{KeldyshJETP65}  \\
effective MPA & $b_K$ (cm$^{2K-3}$W$^{1-K}$) & $8.3\times 10^{-52}$  & $5.4\times 10^{-24}$ & -  \\
critical electron density  & $\rho_c$ (cm$^{-3}$) & $1.7 \times 10^{21}$& $7 \times 10^{21}$ &- \\
plasma absorption & $\sigma_a$ (cm$^{2}$) & $6.3 \times 10^{-18}$& $1.6 \times 10^{-18}$ &Eq. (\ref{eq2}) \\
plasma defocusing & $\sigma_d$ (cm$^{2}$) & $4.4 \times 10^{-17}$& $2.2 \times 10^{-17}$ &Eq. (\ref{eq2}) \\
\hline
linear refractive index & $n_0$ & \multicolumn{2}{|c|}{1.33} & \cite{MielenzAO1978}\\
nonlinear index & $n_2$ (cm$^2$/W) & \multicolumn{2}{|c|}{$1.9 \times 10^{-16}$} & \cite{WilkesAPL09}\\
ionization potential & $U_i$ (eV) & \multicolumn{2}{|c|}{6.5}  & \cite{KennedyI3EJQE95} \\
$\mathrm{e^-}$-ion collision time & $\tau_c$ (fs) & \multicolumn{2}{|c|}{3}  &\cite{KennedyI3EJQE95} \\
recombination rate  & $\eta$ (cm$^3$/fs) & \multicolumn{2}{|c|}{$2\times10^{-9}$} &  \cite{KennedyI3EJQE95} \\
recombination time  & $\tau_r$ (fs) & \multicolumn{2}{|c|}{$100$} &  \cite{MinardiOL09} \\
neutral atom density & $\rho_{nt}$ (cm$^{-3}$) & \multicolumn{2}{|c|}{$6.7 \times 10^{22}$} &  \cite{KennedyI3EJQE95} \\
speed of light in vacuum & $c$ (m/s) &\multicolumn{2}{|c|}{$299792458$}  &-\\
vacuum permittivity & $\epsilon_0$ ($Fm^{-1}$) &\multicolumn{2}{|c|}{$8.85\times10^{-12}$}  & -\\
%electron mass & $m_e (Kg)$ &\multicolumn{2}{|c|}{$9.01\times10^{-31}$}   & -\\
%elementary charge & $e (C)$ &\multicolumn{2}{|c|}{$1.602176565(35)\times10^{-19}$}   & -\\
\hline
\end{tabular}}
\caption{Parameters used in Eqs. (\ref{eq1})-(\ref{eq3}). All numerical results are produced with these values, with the exception of Fig. \ref{f4}(a) in which several values of $b_K$ are used.}
\label{table}
\end{table}

%FIGURE 2
\begin{figure*}
\includegraphics[width=\textwidth]{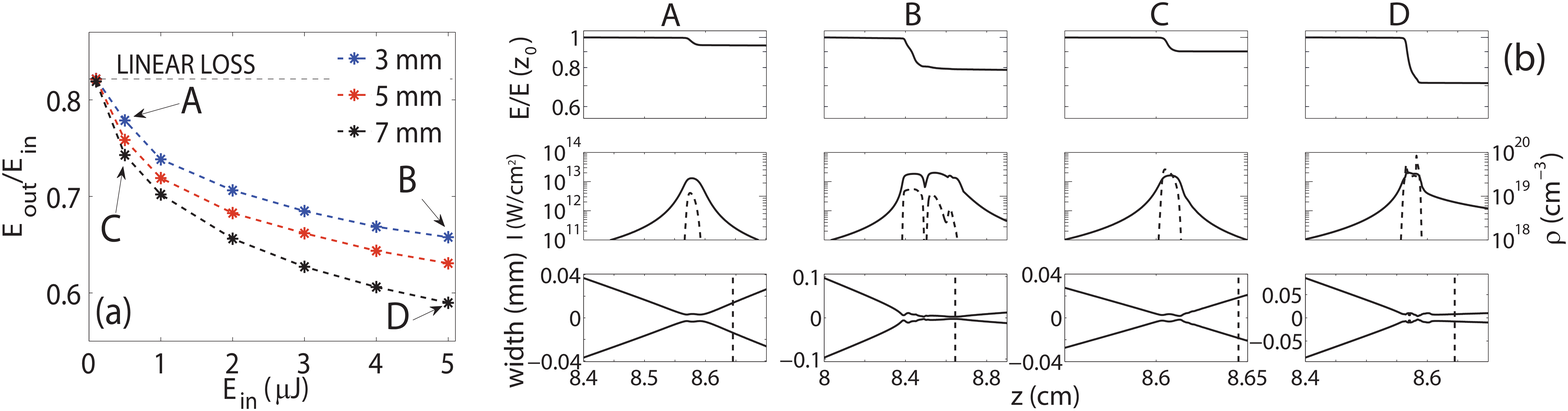}
\caption{(a) Transmission as a function of input energy for several $w_{FWHM}$ (values in legend). The dashed horizontal line marks the transmission in the linear limit, $\exp(-\beta_1L_w)$. (b) Evolution across the nonlinear focus of (top) normalized energy, (center) maximum peak intensity (solid) and maximum plasma density (dashed), and (bottom) maximum beam waist (FWHM of the fluence), corresponding to the points $A-D$ in (a). Vertical lines mark the linear focus, $z_{foc}\approx n_0[d-a]\approx8.64$ cm. \label{f2}}
\end{figure*}

%EQ 2
\begin{equation}
\sigma(\omega) \equiv \frac{\omega_0^2 \tau_c}{c \rho_c} \frac{1}{1-i\omega \tau_c} \label{eq2},
\end{equation}
where $\tau_c$ is related to the mean collision time, $\rho_c\equiv\omega_0^2m_e\epsilon_0/e^2$ is the critical density at which the plasma becomes opaque, $m_e$ the electron mass, and  $e$ the elementary charge. The two terms in $\widetilde{{\cal J}}_{PL}$ associated with the real and imaginary parts of $\sigma(\omega)=\sigma_a+i\sigma_d$ account for plasma absorption and plasma defocusing, respectively. Equation (\ref{eq1}) is coupled to the rate equation \cite{KennedyI3EJQE95,NoackI3EJQE99,CouaironSPR07},

%EQ3
\begin{equation}
\frac{\partial \rho}{\partial t}=W(I) \left[1-\frac{\rho}{\rho_{nt}}\right]+\frac{\sigma_a(\omega_0)}{U_i}\rho I  + \partial_t\rho\vert_{rec}
\label{eq3}.
\end{equation}
From left to right, the terms on the right hand side in Eq. (\ref{eq3}) describe multiphoton ionization (MPI), avalanche ionization (inverse Bremsstrahlung), and electron-ion recombination. Note validity of Eq. (\ref{eq3}) is restricted to the weak plasma condition: $\rho\ll\rho_{nt}$. A list of parameters and their values is provided in table I. The model provided here disregards back reflection of light. Recent numerical studies done with Maxwell solvers on plasma generation in water micro-droplets \cite{EfimenkoJOSAB14} clearly show that plasma induced reflections are of the order of $1\%$ or less. 

The origin of the propagation coordinate, $z$, is chosen at the first air-glass interface of the water tank (see Fig. \ref{f1}). 
At this position, Eq. (\ref{eq1}) is initialized with the pulse

%EQ4
\begin{eqnarray}
&&  {\cal{E}}(t,r,0)=%\\ &&
{\cal{E}}_0\exp\left(-\frac{r^2}{w_{0}^2}\left[1+i\frac{k_0w_{0}^2}{2f_{0}}\right]-\frac{ t^2}{t_{p0}^2}\left[1+iC\right]\right) \label{eq4},
\end{eqnarray} 
where $w_0$, $f_0$, $t_{p0}$ and $C$ denote the input beam width, curvature, pulse duration and chirp, respectively. For comparison with experimental results, we refer below to the beam width $w_{FWHM}$ and pulse duration, $t_p$, at the position of the focusing lens ($z=-a$).
Under our geometrical conditions the generation of plasma is highly localized around the nonlinear focus, which position and size are around $z_{NL}\gtrsim 8.4$ cm and $\Delta z_{NL}\lesssim 0.1$ cm, respectively (see e.g., Figs. \ref{f2}(b) and \ref{f5}), and therefore ${\cal{J}}\approx 0$ in Eq. (\ref{eq1}) during most of the propagation in water. Under these conditions, the beam waist typically decreases from $w_0$ by two or three orders of magnitude before reaching the focal region but the beam and the pulse remain approximately Gaussian. Strong space-time reshaping of the pulse occurs mainly in the focal region. Numerical integration of Eq. (\ref{eq1}) in the region prior to the focus can therefore be advantageously replaced by a less intensive numerical integration of the propagation by using the moment method \cite{AndersonPOF79,BergePOP2000,ZemlyanovAAO05b} up to $z_0\approx8$ cm (see appendix \ref{app} for details).

%Several simulations of Eqs. (\ref{eq1})-(\ref{eq3}) were performed by calculating input conditions just before the nonlinear focus at $z=z_0$ (see appendix \ref{app} for details).% with

This strategy allowed us to perform a parametric study leading to a good match between numerical and experimental data (see Sec. \ref{chirp}), a task that would have been computationally much more expensive with the full model Eqs. (\ref{eq1}-3) due to the fine spatiotemporal resolution required in the numerical grids  to convey an input $7$ mm wide beam through the focus. We checked with benchmarks that results are not significantly affected by the use of the moment method in the first propagation stage.

% SEC IV
\section{nonlinear energy deposition and plasma excitation at $\lambda_0=800$ \MakeLowercase{nm}}

In section \ref{focgeom} we review the numerical results regarding the effects on energy deposition induced by changes in the focusing geometry. We then show in Sec. \ref{chirp} that further tuning of the transmission and plasma volumes is achieved simply by modifying the input pulse chirp, in the geometry used in experiments.

% SEC IV A
\subsection{Influence of beam width and pulse energy}
\label{focgeom}

Figure \ref{f2}(a) shows the optical transmission obtained for different focusing geometries, i.e., varying $w_{FWHM}$ from $3$ to $7$ mm and $E_{in}$ from $0.1$ to $5$ $\mu$J (pulses are initially un-chirped: $C=0$, $t_p=t_{pl}=45$ fs). The drop in transmission when increasing either $E_{in}$ or $w_{FWHM}$ is due to the enhancement of multiphoton and/or plasma absorption. This happens because the localization of light imposed by the lens and further enhanced by self-focusing on the transverse plane, is more efficient than all other processes that yield decrease of intensity, such as pulse dispersion \cite{PotasekJOSAB86} and plasma defocusing \cite{KiranPRA10,KiranOE10}.

Figure \ref{f2}(b) shows the evolution of energy, peak intensity and plasma, and beam waist across the nonlinear focus for four different situations ($A$, $B$, $C$, and $D$ in Fig. \ref{f2}(a)). Whilst low energy deposition (columns $A$, $C$) is associated with a short focal region of high intensity and a nearly symmetric diffraction from the focal plane, the situations in which high energy is deposited to the water ($B$, $D$) present a high intensity region showing one (or several) flat plateaus. In the latter cases the beam does not diffract as fast as in the former ones due to the prolongated effect of self-focusing, which keeps the beam waist relatively narrow despite the presence of plasma. Indeed, the range of parameters in the case $D$ suggests it falls in the filamentation regime \cite{LiuAPB03}.

\subsection{Influence of the input pulse chirp}
\label{chirp}

%FIGURE 3
\begin{figure}
\includegraphics[width=0.5\textwidth]{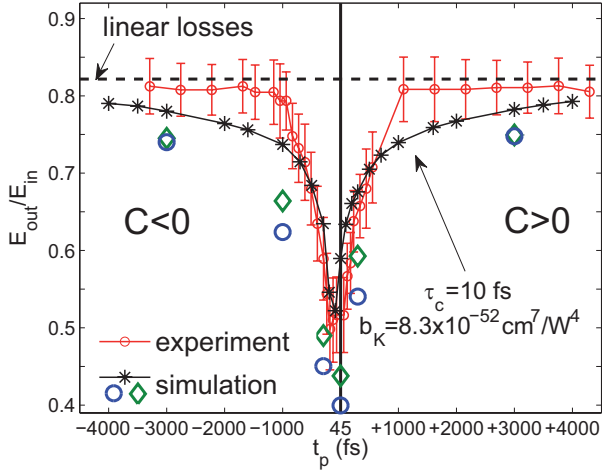}
\caption{Numerical and experimental transmission along the $L_w=10$ cm of water as a function of the input pulse width at $\lambda_0=800$ nm. Numerical data in circles and diamonds show results obtained when full Keldysh ionization plus quadratic (dashed) or linear (dotted-dashed) electron-ion recombination are introduced in Eqns. (1)-(3). The solid vertical line marks the minimum pulse width, $t_{pl}=45$ fs, and the sign of the other $t_p$ values denotes the sign of the chirp, C. Linear transmission is marked by the horizontal line. Experimental curve corresponds to post processed \textit{raw} data accounting for the Fresnel reflections at the air-glass and glass-water interfaces as to reflect only the transmission inside the water tank, as in numerical results.
Error bars correspond to the standard deviation obtained over $11$ measurements.
\label{f3}}
\end{figure}

%FIGURE 4
\begin{figure}
\includegraphics[width=0.5\textwidth]{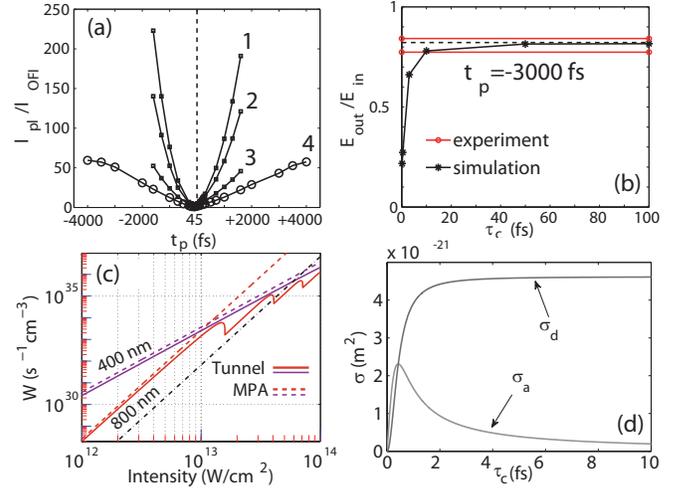}
%\includegraphics[scale=.27]{LOSSpl2mpa_Shanghai.eps}
%\includegraphics[scale=.27]{Etrans_vs_tc_Shanghai.eps}
%\\
%\includegraphics[scale=.27]{KeldyshRATEvsI.eps}
%\includegraphics[scale=.27]{Sigma_av_and_def.eps}
\caption{(a) Plasma absorption to MPA ratio for several values of $b_K$ and $\tau_c$ as a function of input pulse width: (1) $b_K=3.6\times10^{-51}$ cm$^7$/W$^4$, $\tau_c=3$ fs; (2) $b_K=6.2\times10^{-51}$ cm$^7$/W$^4$, $\tau_c=3$ fs; (3) $b_K=\beta_K=3.6\times10^{-50}$ cm$^7$/W$^4$, $\tau_c=3$ fs; (4) $b_K=8.3\times10^{-52}$ cm$^7$/W$^4$, $\tau_c=10$ fs. Vertical line marks the minimum pulse duration. (b) Energy deposition as a function of $\tau_c$ for fixed $t_p$ ($b_K=8.3\times10^{-52}$ cm$^7$W$^{-4}$). Solid horizontal lines delimit the experimental interval of the measurement (see Fig. \ref{f3}) and the dashed one the linear losses. (c) Keldysh ionization rate, $W$,  (solid) versus $I$ for water at $\lambda_0=800$ and $400$ nm. Dashed lines account for MPI only and the dashed-dotted shows our effective rate. (d) plasma absorption and defocusing dependence on collision time.\label{f4}}
\end{figure}

The effect of input pulse chirp, $C=\pm[\{t_p/t_{pl}\}^2-1]^{1/2}$, was studied for $E_{in}=5\ \mu$J and $w_{FWHM}=7$ mm, corresponding to the experimental conditions (the relatively wide beam is required to keep intensity below the BK7 glass damage threshold \cite{TzortzakisPRL01b}). The input pulse width was varied from $t_{p}=45$ fs ($C=0$) to $t_{p}\approx4$ ps, for positive and negative chirp. Figure \ref{f3} shows the transmission as a function of the input pulse width obtained numerically, by integrating Eqs. (\ref{eq1})-(\ref{eq3}) (stars), and experimentally, in the setup of Fig. \ref{f1} (circles). Note all input pulses have exactly the same bandwidth. Moreover, because we are in the deep normal GVD regime of water self phase modulation (SPM) is the main frequency conversion effect which does not widen substantially the spectra at the energy levels and focusing conditions used here. Therefore, all results presented below are interpreted solely in terms of the spatiotemporal dynamics of the different pulses and spectral broadening effects are dismissed in all discussions (see, e.g., Ref. \cite{VasaPRA14} for recent results on supercontinuum generation in water by pumping close to the zero GVD, and Refs. \cite{DudleyRMP06,SkryabinRMP10} for reviews on the topic in one-dimensional systems).

Experimental results in Fig. 3 present a minimum transmission for negatively chirped pulses ($t_p\approx-100,-150$ fs, the minus sign stands for negative chirp). This is expected because water exhibits normal GVD at $\lambda\sim800$ nm, and therefore the maximum intensity levels at focus (maximizing nonlinear losses) will be achieved when the pulses reach this region with nearly compensated chirp $C(z_{NL})\approx0$. For highly chirped pulses the levels of transmission tend $\sim0.82$, in excellent agreement with the predicted value in the linear regime, $\exp(-\beta_1 L_w)\approx0.82$.

%FIGURE 5
\begin{figure*}
\includegraphics[width=\textwidth]{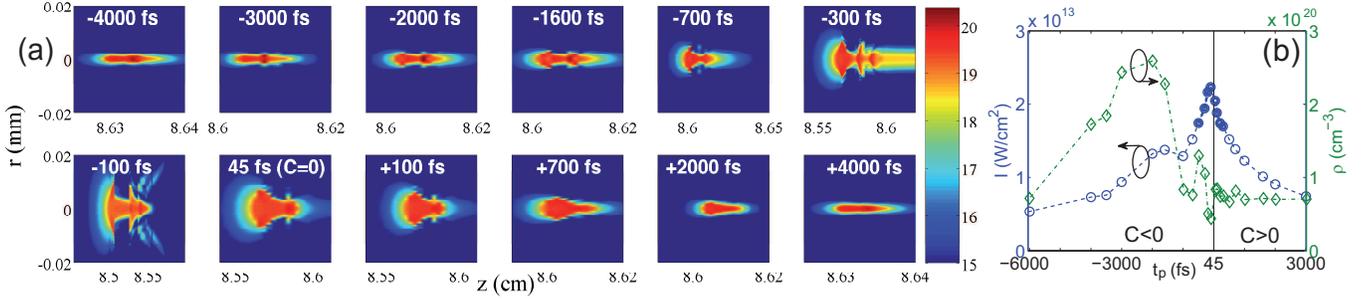}
\caption{(a) Plasma volumes around the focus for several initial pulse widths (marked by the labels). Densities are given by $\rho(r,z)=10^x$ cm$^{-3}$, where the power $x$ is given by the color bar. (b) absolute maxima of the electron density and intensity achieved in the focal volumes as a function of the chirped input pulse widths. \label{f5}}
\end{figure*}

%FIGURE 6
\begin{figure}
\includegraphics[width=0.5\textwidth]{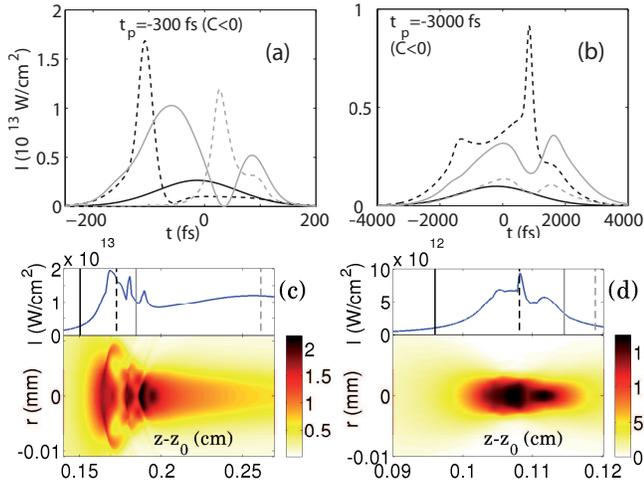}
\caption{Temporal intensity and plasma profiles at $r=0$ for several distances and input widths, $t_p$, of (a) $-300$ fs and (b) $-3$ ps. (c, d) Evolution across the nonlinear focus of (top) maximum of intensity and (bottom) fluence in J/cm$^2$. The color and style of the lines link the temporal profiles (a,b) with the distances marked by the vertical lines in (c,d), respectively. $z_0=8.4$ ($8.5$) cm for left (right) column. \label{f6}}
\end{figure}

Numerical simulations were carried out by solving Eqns. (1)-(3). Fitting the experimental results involved a two parameter study in which the OFI rate, $W(I)$, and the collision time, $\tau_c$, were tuned. The reason for this was that by using the Keldysh theory it was not possible to match the experimental data within the error bars. In Fig. 3, diamonds and circles show numerical data obtained with Keldysh theory, Drude model with $\tau_c=10$ fs, and recombination terms which coefficients are of the order of the empirically determined ones: $\partial_t\rho\vert_{rec}=-\eta\rho^2$, $-\rho/\tau_r$, respectively  (see Table I). Results are systematically $\sim10\%$ below the measured transmission and a further increase of $\tau_c>10$ fs did not improve substantially this picture (note plasma absorption cross section, $\sigma_a$, is already close to zero). In principle it is possible to tune certain parameters in the Keldysh model (developed for solids, rather than liquids), such as the ionization potential or the electron-hole mass ratio which contains information of the actual curvature of the valence and conduction bands. However, these features of water remain unknown to date (see, e.g., Refs. \cite{BernasJPPA98,MinardiarXiv14} for a discussion of the different measurements of the ionization potential in water and how these help matching numerical and experimental data). For fitting purposes, we then opted for tuning the ionization rate through an effective parameter $b_K$: $W(I)\approx b_KI^K/U_i$, and electron-ion recombination $\sim\rho/\tau_r$. The reason why the use of this OFI model in Eq. (1) is justified is discussed below. %The relative importance of OFI and plasma absorption in the overall transmission.
In the slowly varying envelope model Eqs. (\ref{eq1})-(\ref{eq3}), the instantaneous optical \textit{kinetic} power density transferred to the medium due to OFI and plasma absorption is given by (see, e.g., \cite{Haus_book_2000})

%EQ 5
\begin{eqnarray}
&& \nonumber 
{\cal{W}}(t,r,z)\equiv\frac{1}{2}\mathrm{Re}\left\{{JE^*}\right\}=
W(I)U_i\left[1-\frac{\rho}{\rho_{nt}}\right]+{\sigma_a}\rho I,
\\ &&\label {eq8}
\end{eqnarray}
and the loss ratio $l_{pl}/l_{OFI}$, shown in Fig. 4(a), is calculated from:

%Figure \ref{f4}(a) shows the ratio of losses induced by plasma and OFI, $l_{pl}/l_{OFI}$: 

%EQ 6
\begin{eqnarray}
&& %\nonumber 
l_{pl}\equiv\left.\int rdrdzdt\ {\cal{W}}\right|_{W(I)=0},%\\ &&
\ l_{OFI}\equiv\left.\int rdrdzdt\ {\cal{W}}\right|_{\sigma_a=0}\label{eq9}.
\end{eqnarray}
Whilst avalanche ionization governs nonlinear losses for large values of chirp, OFI acquires its maximum importance around $t_p\approx-100,-150$ fs, where transmission is minimized (see Fig. \ref{f3}). Note from Fig. 4(a) that this is rather general for the relatively wide range of $b_K$, $\tau_c$ values. Therefore, in first place $\tau_c$ was chosen as to match experimental data (within the error bars) in the transmission asymptotes, $|t_p|\gtrsim3000$ fs. Figure \ref{f4}(b) shows the typical effect the collision time, $\tau_c$, has on the transmission for large input chirp, $C$. Large values of collision time, $\tau_c\gg10$ fs, tend to remove plasma absorption (see Fig. \ref{f4}(d)) and the transmission tends to the linear one. This highlights the fact that $OFI$ is much less important than avalanche ionization for large values of input pulse chirp. In second place, $b_K$ was adjusted to match experiments around the minimum of transmission, $t_p\approx-100,-150$ fs. The resulting pair of parameters ($b_K$, $\tau_c$) obtained by this method (see table \ref{table} or Fig. \ref{f3}) provided a good quantitative agreement with experimental results.

In the situations where OFI maximizes its effect, the maximum intensity levels typically fall in the range $1.6\times10^{13}\lesssim I_{max} \lesssim2.2\times10^{13}$ W/cm$^2$ (see Fig. 5(b)), and therefore $\log{W}$ vs $\log{I}$ follows a linear trend, as shown in Fig. 4(c). Since the corresponding slope is very similar to the one of the pure multi photon absorption (MPA) regime (see dashed line in Fig. 4(c)), the OFI term proposed above, $W(I)\approx b_KI^K/U_i$, is justified and $b_K$ plays the role of an \textit{effective} MPA parameter. Accordingly, the $b_K$ value providing a good match is an order of magnitude smaller than $\beta_K$ (see Table 1): $b_K/\beta_K\approx W(I_{max})/W(0)\sim10^{-1}$.

%The reason whiy $b_K$ is an order of magnitude smaller than $\beta_K$ (see Table 1) is because the ionization rate does not fall in the pure MPI regime for the typical intensity levels involved here, as it can be seen from Figs. 4(c) and 5(b). In Fig. 5(b), the intensity levels above the first Keldysh rate fall (at $I\approx1.6\times10^{13}$ W/cm$^2$) are marked by the thick trace. For these high intensities, the linear trend of $\log{W}$ vs $\log{I}$ implies that using an \textit{effective} OFI coefficient, $b_K$, in Eqs. (\ref{eq1}-\ref{eq3}) is equivalent as to the full Keldysh rate. It could be argued that this effective value of $b_K$ does not apply for the lower intensity situations, for which $I_{max}<1.6\times10^{13}$ W/cm$^2$ ($E_{out}/E_{in}\gtrsim0.7$ in Fig. \ref{f3}) because it underestimates OFI losses. However, as discussed above, nonlinear losses are typically dominated by plasma absorption in this region by one order of magnitude or more and OFI is responsible for only a few percent of the total losses. Note that we are not fitting here the Keldysh rate, but an effective MPA coefficient that mimics the Keldysh rate at the intensity levels of interest, i.e., those for which OFI induced losses acquire their maximum importance.

%FIG 7
\begin{figure*}
\includegraphics[width=\textwidth]{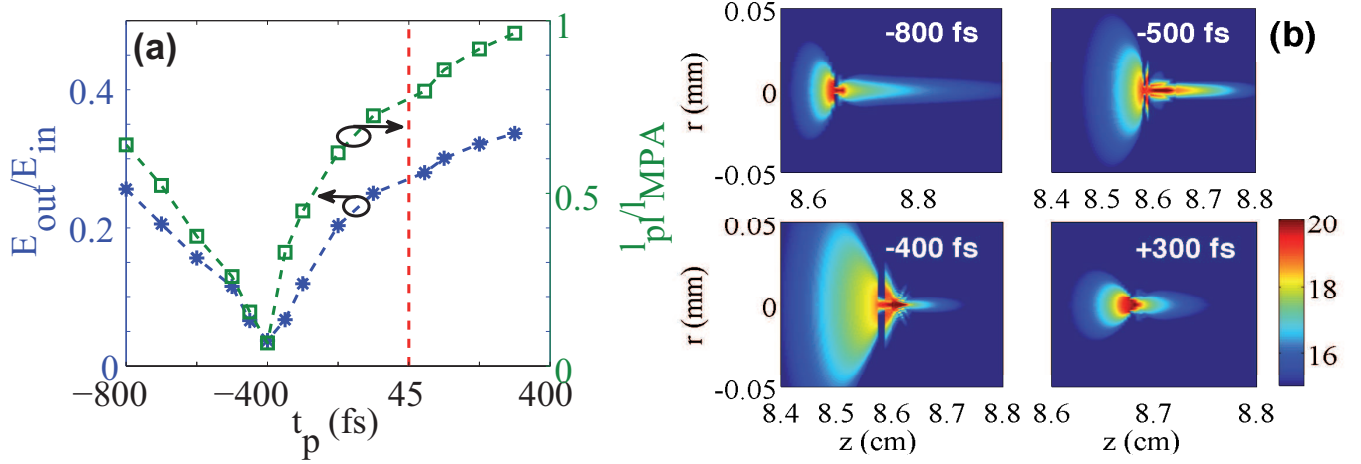}
\caption{(a) Transmission (left) and ratio of plasma absorption to OFI (right) as a function of the input pulse duration. Dashed vertical line marks the minimum pulse width ($C=0$). (b) Plasma volumes for several input pulse widths (see labels) and electron densities $\rho=10^x$ cm$^{-3}$, where $x$ is given by the color bar. \label{f7}}
\end{figure*}

Figure \ref{f5} shows maxima of the electron plasma density and intensity around the nonlinear focus for several input pulse widths, $t_p$. Comparison with Fig. 3 reveals several interesting features. First, we note that the minimum of absorption in Fig. 3 corresponds exactly with the case in which the maximum intensity is achieved (see Fig. 5(b)). This might thought to be expected because indeed OFI seeds more free electrons at higher intensities and avalanche effects induce further losses to the optical field. However, the overall absorption is a spatiotemporal integrated quantity whereas the maximum intensity represents only a local feature. The fact that there is a correspondence in between the local and integrated quantities highlights the importance of strong OFI in the overall losses for locally high intensity levels, since it is the highest order nonlinear effect ($\sim I^5$). For this value of the input chirp, $t_p=-150$ fs, the plasma volume having absorbed the biggest possible energy is generated (note this is relatively easy to identify in experiments).

By further negatively pre-chirping the input pulses to $t_p=-300$ fs we obtain the longest possible plasma volume, shown in Fig. 5(a). This is physically understood from Figs. \ref{f6}(a) and (c) showing several temporal profiles around the focal region, maxima of intensity, and fluence. The pulse reaches the focal volume with $t_p(z_{NL})\approx-150$ fs (see the solid black temporal profile in Fig. \ref{f6}(a), corresponding to $z-z_0=0.15$ cm in Fig. 6(c)). In this case, intensity is high enough for the Kerr effect to produce the refocusing cycles observed in the fluence, Fig. \ref{f6}(c), that keep the on axis intensity relatively high, $\gtrsim10^{13}$ W/cm$^2$, for $\sim1$ mm (at the level of $\rho\sim10^{17}$ cm$^{-3}$). Indeed, refocusing cycles are linked to the local increase of intensity in the temporal profiles and to the long high intensity and high plasma density region. Note, interestingly, that the maximum plasma channel length does not coincide exactly with the $t_p$ minimizing transmission ($t_p\approx-100$, $-150$ fs), but, instead, the input pulse has to be slightly further negatively chirped ($t_p\approx-300$ fs). This is simply because in the latter case, the pulse still undergoes temporal compression on its way through the focus, which helps keeping higher intensities.

% further pre-chirp shortens volume and enhances plasma: effect of plasma absopriton (dominant!)

Transmission values in Fig. 3 rapidly increase towards the linear transmission limit  for pulse widths $t_p\lesssim-700$, $\gtrsim45$ fs. In these cases the plasma tracks are dramatically shortened down to $\sim100$ $\mu$m. However, on the negative chirp side, the maximum plasma density achieved presents an increasing trend  as input duration increases up to, $t_p\sim-2000$ fs, as shown in Fig. 5(b). For longer input pulses, maximum of plasma density drops, as expected for very long pulses with finite energy. The fact that the maximum plasma density is located for relatively strongly chirped pulses ($t_p\approx-2$ ps) highlights the increasing importance of avalanche effects with $t_p$ (see Fig. 4(a)). Interestingly, such density increase observed in Fig. 5(b) is associated to the existence of a hump in intensity at $t_p=-1500, -2000$, shown also in Fig 5(b). This hump is originated from small refocusing cycles that are manifested as intensity spikes after the first focusing event. This is caused by a pronounced and peaked trailing part of the pulse (after the leading one is attenuated) reaching the maximum intensity of the whole pulse propagation. This intense trailing part further accelerates electrons and is therefore able to further increase plasma density. Note that this feature of the temporal profile is shown in Fig. 6(b): the dashed black profile there is precisely the one corresponding to the intensity spike seen in the top part of Fig 6(d). This effect is possible only if the back part of the pulse is time-shifted from the front part by a delay smaller than recombination times. Outside the intensity hump, the maximum intensity is always reached by the leading part of the pulse. This explains the often experimentally reported observation that at a constant pulse energy, the longer the incident pulse is, the more permanent damage is produced in solids (see Ref. \cite{OndaJOSAB05} for similar experiments in synthesized silica). When these long pulse reach the focus their low peak intensity has the double drawback that less plasma is excited and self-focusing is weaker, therefore high intensity region is shorter and less intense, as shown in Figs. \ref{f6}(b) and \ref{f6}(d).

Summarizing the results presented in this section, depending on the desired application, plasmas may be substantially long for the slightly negative chirped pulses at focus, or plasmas may be shortened and highly localized with higher densities for pulses pre-chirped up to a few ps.

%SEC V
\section{Results for $\lambda_0=400$ \MakeLowercase{nm}}

A motivation to study the effects presented in the above sections at the shorter $\lambda_0=400$ nm is that linear losses of water are close to their minimum for this wavelength (see $\beta_1$ in table \ref{table}), so the focal point can be placed, in principle, at much longer distances from the laser source ($\sim50$ times farther with same attenuation). However, in this case ionization will set up at lower intensities than for $800$ nm (see Fig.  \ref{f4}(c)), providing seed electrons for avalanche effects at earlier stages relative to the focus. Since plasma cross section is almost constant for these wavelengths the total nonlinear losses, $\int rdrdtdz{\cal{W}}$, will be enhanced (intensity levels were observed to stay relatively similar in both cases). This is in agreement with numerical findings, Fig. \ref{f7}(a), where the minimum of transmission requires bigger pre-chirping than in Fig. \ref{f3} simply because normal GVD is stronger at $400$ nm. The increase in the ionization rate is such that nonlinear losses are now dominated by OFI, as shown in Fig. \ref{f7}(a). Intensity levels achieved here ($\sim10^{13}$ W/cm$^2$) suggest that $b_K$ would not need to be significantly modified from the Keldysh value to match experimental results since the rate is dominated by OFI in this case (see Fig.  \ref{f4}(c)), oppositely to the $800$ nm case.

Plasmas obtained here are shown in Fig. \ref{f7}(b) and share all qualitative features with those observed at $800$ nm, i.e., elongation close to the minimum of the optical transmission and shortening otherwise, suggesting these properties are rather general and wavelength independent.

%CONCLUSIONS
\section{conclusions}

We showed that the control on focusing conditions and input pulse chirp provides a simple and effective mechanism to modify at will the electron-plasma density distribution generated by high energy ($\sim5\ \mu$J) fs-pulses in water. Pulses reaching the focal volume with the shortest temporal profiles generate elongated plasma regions with relatively constant high densities. Plasma densities can be made higher by further pre-chirping the pulses (up to few ps), at the expense of the plasma channel length. Nonlinear losses, dominated by plasma absorption at $800$ nm and by OFI at $400$ nm, are found to be maximized for the cases in which OFI acquires its maximum efficiency. These remarks are independent of wavelength across the visible spectrum. The different plasma density distributions provide an idea of where the laser energy is deposited by the pulse and the relative amount of it. This knowledge might be useful for developing applications of laser energy deposition to medical therapies and surgeries as well as to the controlled generation of sound waves in water.

\section{acknowledgments}
Authors acknowledge financial support from the French Direction G\'{e}n\'{e}rale de l'Armement (DGA).

%APPENDIX
\appendix
\section{Numerical model far from NL focus}
\label{app}

The need of a model that describes spatiotemporal propagation far from nonlinear focus is easily justified as follows. Under our experimental conditions, an order of magnitude estimate of the beam waist at focus, $w_f$, may be obtained from initial (at position of the lens, $z=-a$) waist, $w_i=w_{FWHM}/\sqrt{2\ln2}$, curvature, $f_i$, and $a$, $d$ (lens separation from water tank and focal length), with the laws of Gaussian (linear) optics  \cite{Svelto2010} $f=\xi+Z_{min}^2/\xi$, $w^2=w_{min}^2[1+\xi^2/Z_{min}^2]$, where $w_{min}$ is the minimum waist and $Z_{min}\equiv\pi w_{min}^2/\lambda_0$ the associated Rayleigh length in vacuum:

\begin{eqnarray}
&& w_f^2=\frac{\lambda_0}{2\pi n_0}\tilde{Z}_{a}\left[1-\sqrt{1-\frac{4[d-a]^2}{\tilde{Z}_{a}^2}}\right]\label{eqA1},\\ && \nonumber
\tilde{Z}_{a}\equiv n_0Z_a=n_0\frac{\pi w_{a}^2}{\lambda_0}=n_0\left[Z_f+\frac{\{d-a\}^2}{Z_f}\right],\\ &&  Z_f=\frac{d\lambda_0}{\pi w_i^2}f_i,\ \nonumber
f_i=\frac{Z_i^2}{2d}\left[1-\sqrt{1-\frac{4d^2}{Z_i^2}}\right],\ Z_i\equiv\frac{\pi w_i^2}{\lambda_0},
\end{eqnarray}
where $w_{a}$ and $\tilde{Z}_{a}$ are the beam waist and Rayleigh length at $z=0^+$ cm, i.e., just inside the water tank, respectively. In this situation, for the $w_{FWHM}=7$ mm beam we typically have $w_{a}/w_f\approx10^3$ and $Z_f\approx30\ \mu$m (numerical aperture $NA\equiv w_{FWHM}/[2n_0d]\sim0.03$). This implies that the computational grids discretizing $r$, $z$ should contain $\sim10^5$, $10^4$ points, if the window widths are to be of the order $2\Delta r\sim5w_{a}$, $2\Delta z\sim 10L_w/Z_f$, to properly resolve the dynamics through the focal point, computation times taking up to weeks, exceeding any practical time scale, particularly in the frame of a parametric study demanding many simulation runs.

In our simulations, the typical input intensities at the entrance of the water tank, $I\sim10^8$ W/cm$^2$, are well below ionization thresholds ($J\approx0$ in Eq. (\ref{eq1})). Pulse powers, however, are larger than the critical power for self-focusing, hence, the Kerr term dominates nonlinear effects until the increase of intensity due to beam focusing is such that typical Kerr and MPA lengths satisfy $L_{Kerr}\equiv[n_2k_0I]^{-1}\ll L_{MPA}\equiv[2\beta_KI^{K-1}]^{-1}$ if $I\ll [n_2k_0/\left\{2\beta_K\right\}]^{1/[K-2]}\sim10^{12}$ W/cm$^2$, in our range of wavelengths. Below this intensity level, it is a reasonable approximation to assume that the propagation of the input beam ($\Psi\equiv{\cal{E}}/\sqrt{\epsilon_0cn_0/2}$: $I=\vert \Psi\vert^2$)

\begin{equation}
\Psi(t,r,z=0)=\Psi_0\exp\left(-\frac{r^2}{w_i^2}\left[1+i\frac{k_0w_i^2}{2f_i}\right]-\frac{ t^2}{t_{p0}^2}\left[1+iC\right]\right)\label{eqA2},
\end{equation}
can be described accurately by the method of pulse characteristics (see e.g., \cite{ZemlyanovAAO05b}), which describes the evolution in terms of the pulse moments. We define pulse energy, square beam radius and pulse duration as

\begin{subequations}
\begin{eqnarray}
&& U(z)=2\pi\int_{V} |\Psi(t,r,z)|^2 \label{eqA3},\\ &&
R^2(z)=\frac{2\pi}{U(z)}\int_{V} r^2|\Psi(t,r,z)|^2,\\&&
T^2(z)=\frac{2\pi}{U(z)}\int_{V} t^2|\Psi(t,r,z)|^2,
\end{eqnarray}
\end{subequations}
where $\int_V\equiv\int_0^{\infty} r\mathrm{d}r\int_{-\infty}^{\infty}\mathrm{d}t$. Evolution equations for the above quantities are obtained through moment \cite{BergePOP2000} or variational \cite{AndersonPOF79}  methods that reduce the dimensionality of the problem by transforming Eq. (\ref{eq1}) with $\mathbf{J}=\mathbf{0}$ to (dots denote $z$-derivatives: $\dot{x}\equiv\mathrm{d}x/\mathrm{d}z$):

\begin{subequations}
\begin{eqnarray}
\dot{U}&=&-\beta_1 U \label{eqA4},\\ 
\ddot{R^2}&=&\frac{2\pi}{k_0^2 U}\left[ \int_{V} |\nabla_{{\mathbf r}} \Psi|^2 - \frac{4\pi}{\mathscr{P}_{cr}}\int_{V} |\Psi|^4  \right],\\
\ddot{T^2}&=&\frac{4\pi k_0''}{U}\left[k_0'' \int_{V} |\partial_{t} \Psi|^2 + \frac{2\pi}{k_0\mathscr{P}_{cr}}\int_{V} |\Psi|^4 \right],
\end{eqnarray}
\end{subequations}
where $\mathscr{P}_{cr}\equiv\alpha\lambda_0^2/8\pi n_0n_2$ is the critical power for beam collapse ($\alpha=4$ in this analytical approach) and the pulse evolution is assumed to remain Gaussian:

\begin{multline}
\Psi(t,r,z)=\Psi_0(z)\exp\left(-\frac{r^2}{2R^2(z)}\left[1+i{k_0R(z)\dot{R}(z)}\right] \right ) \\
\times
\exp\left(-\frac{ t^2}{4T^2}\left[1-i{T(z)\dot{T}(z)/k_0"}\right]\right)\label{eqA5}.
\end{multline} 
Inserting Eq. (\ref{eqA5}) into Eqs. (\ref{eqA4}), we eliminate the pulse intensity from the relation $U(z)=\pi^{3/2}\sqrt{2}R^2(z)T(z) \Psi_0^2(z)$ and find,

\begin{subequations}
\begin{eqnarray}
\dot{U}&=&-\beta_1 U \\ 
\ddot{R}&=&\frac{1}{k_0^2 R^3} \left[1-\frac{U}{2\sqrt{\pi} T \mathscr{P}_{cr}}\right]\\
\ddot{T}&=&\frac{k_0''^2}{4T^3}\left[1+\frac{U T}{2\sqrt{\pi} k_0 k_0'' \mathscr{P}_{cr}R^2}\right],      
\end{eqnarray}
\label{eqA6}
\end{subequations}
which is to be solved with the initial conditions $U(0)=E_{in}=\mathscr{P}_{in}t_p\sqrt{\pi/2}$, $R(0)=w_0/\sqrt{2}$, $\dot{R}(0)=-w_0/\sqrt{2}f$, $T(0)=t_p/2$ and $\dot{T}(0)=2k_0''C/t_p$, where the chirp $C=\pm[t_p^2/t_{pl}^{2}-1]^{1/2}$, $t_{pl}=45$ fs being the width of the transform limited pulse as emitted by the laser source.

We use Eqs. (\ref{eqA6}) to estimate the initial conditions for Eq. (\ref{f1}): $w(z_0)$, $f(z_0)$, and $t_{p}(z_0)$ inside the water tank for which $I\approx10^{11}$ W/cm$^2$. For our input energies $E_{in}\leq5\ \mu$J, this method lets us approach significantly the focal region, in the sense that $w(z_0)\sim10 w_f$.

%% FIRST DO THIS, THEN COPY .BBL CONTENT AND (go to *)
%\bibliography{arnaud0fil}
%\bibliographystyle{osajnl2}

\newcommand{\noopsort}[1]{} \newcommand{\printfirst}[2]{#1}
 \newcommand{\singleletter}[1]{#1} \newcommand{\switchargs}[2]{#2#1}

% * % PASTE HERE AND UNCOMMENT LINES BELOW
%\begin{thebibliography}{99}
%% Do not include separate BibTeX files; if BibTeX is used,
%% paste the output (contents of .bbl file) here.
%\bibitem{revtex-au} \url{https://authors.aps.org/revtex4/}.
%\bibitem{osastyle} \url{http://www.opticsinfobase.org/submit/style/jrnls_style.cfm}.
%\end{thebibliography}

\end{document}